# Situational Awareness with PMUs and SCADA

HyungSeon Oh, *Member, IEEE*

*Abstract*--Phasor measurement units (PMUs) are integrated to the transmission networks under the smart grid umbrella. The observability of PMUs is geographically limited due to their high cost in integration. The measurements of PMUs can be complemented by those from widely installed supervisory control and data acquisition (SCADA) to enhance the situational awareness. This paper proposes a new state estimation method that simultaneously integrate both measurements, and show an outstanding performance.

*Index Terms*--chaos, phasor measurement units (PMUs), state estimation, supervisory control and data acquisition (SCADA).

## I. Nomenclature

| | |
|---|---|
| $d$ | Lyapunov dimension |
| $e_j$ | $j^{th}$ column vector of an identity matrix |
| $H$ | Heaviside function |
| $N$ | number of buses |
| $n_{min}$ | ratio between auto-correlation time and time delay in the measurements |
| $P$ | perfect permutation matrix |
| $\tau_L$ | Lyapunov exponent |
| $v_x$ | real component of voltage vector |
| $v_y$ | imaginary component of voltage vector |
| $v$ | complex voltage vector, $v = [v_x; v_y]$ |

## II. Introduction

IN the electric power system operation, a central dispatcher collects the information on generators' offers and demand forecast, and determines the generation dispatches. However, the modern power systems face fundamental challenges with increased uncertainties. To mitigate the impact of the increased uncertainties, an unplanned event needs to be recognized before the event leads to damages on the power system. State estimation plays a central role in monitoring for reliable control power systems. Power flows and voltage magnitudes are measured through asynchronous SCADA (supervisory control and data acquisition) systems every 2–3 seconds. PMUs (Phasor measurement units) are integrated to enhance the real-time situational awareness under the smart grid umbrella [1]. PMUs provide synchronized direct measure of the voltages at 60Hz, and they are integrated in the U.S. grid systems in 2016 with total funding of $358 million [2]. Less than 1% of all the substations in the transmission grids in the United States are equipped with PMUs, of which 1,400 are integrated to the transmission networks. More importantly, 1/3 of the North American continent has a limited coverage from the PMUs [21]. Therefore, the PMUs are insufficient to directly measure all the voltages. Therefore, data from PMUs are not sufficient to directly measure all voltages over the entire network. Research shows that approximately 30% of PMU integration allows an entire visibility to the grids [3], but it is not practical to achieve the level of integration due to the high costs associated with PMUs. While SCADA covers the entire area, the data is also limited in terms of time. Even though either PMUs or SCADA are not sufficient for situational awareness due to these geographical and temporal limitations, they can be complementary when combined. It would be ideal to estimate the state using data from both SCADA and PMUs.

However, there are fundamental difficulties to do so with traditional state estimators: 1) the large amount of data makes it difficult to detect events of interest and analyze them for power flow studies, while the convergences of the traditional tools are not fast enough for real-time monitoring; and 2) the limited capability in the tools for power flow analysis. The Newton-Raphson method [4] is most frequently used in the state estimation, which ignores the second or higher order terms from the Taylor series expansion [5]. This approach is reasonable if the state variable is well-defined (i.e., all the measurements are corresponding to a single point). However, due to different PMU and SCADA measurement frequencies, the assumption may not hold, which makes the system estimation imprecise when only PMUs are updated and combined with the old data from SCADA. Therefore, the Taylor series after ignoring the higher order terms do not provide an acceptable approximation of the original power balance equations when both measurements are taken at different states. Other power flow methods such as Gauss-Seidel [5] and Holomorphic embedding [6] take a very long time for the traditional power flow algorithms to converge, and in some cases, they may converge a wrong estimate with a very large error, i.e., they are more prone to errors in the same circumstance. Therefore, no studies take both data into consideration simultaneously.

This paper aims to address *situational awareness* from the detection of an anomaly in the power system operation and *precise state estimation* based on tensor-computation to combine data from SCADA and PMUs. Fig. 1 illustrates how the proposed state estimation identifies the relevant data to estimate the post-event state.

This work was supported in part by the U.S. Department of Energy under Grant DE-AC02-05CH11231.

H. Oh is with the State University of New York at Buffalo, Buffalo, New York, 14260 USA (e-mail: hso1@buffalo.edu).

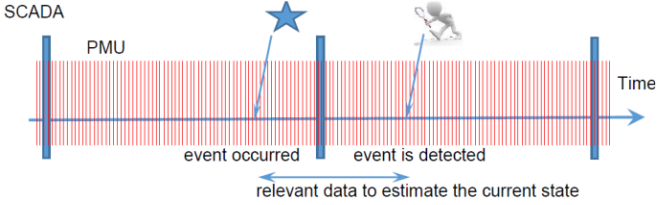

Figure 1. Schematic showing how the proposed work identifies relevant data to estimate post-event states.

## III. LITERATURE REVIEW

### A. System Monitoring with PMUs

For real-time monitoring, four signal processing methods were suggested—fast Fourier transformation, matrix-pencil method, Yule-Walker spectral method, and min-max method—and tested on data from the Texas Synchrophasor Network [7]. By combining the results from the four methods, it is advised to alert an event when three methods report an outlier. They successfully identify an event caused by sudden losses of power above 450 MW. The results from the four signal processing methods are illustrated in Fig. 2 and each method identifies different outliers for the same time series measurements.

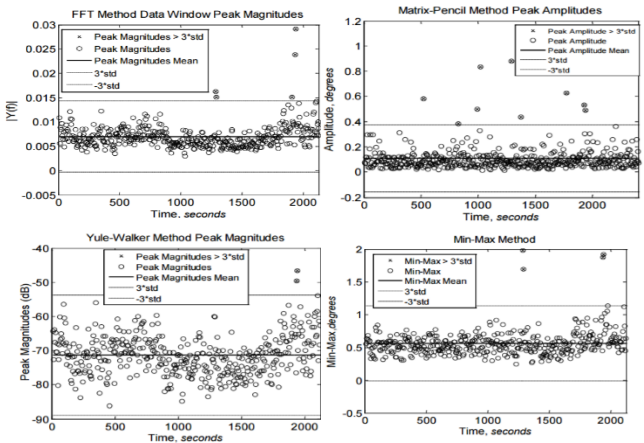

Figure 2. The performance of signal processing techniques applied to a PMU data set [3]; fast Fourier transformation (top left); matrix-pencil method (top right); Yule-Walker spectral method (bottom left); and min-max method (bottom right).

The computation time involved with the methods can be large if checking all four. More importantly, alerting from three methods does not guarantee that there will not be a false alarm, and slow but consistent change such as slow increase in wind generation can be missed in this alerting system (missing event). Situational awareness detects an event that is not known in advance. An ideal alert should be data-driven (not model dependent), guarantee it will detect an event if one occurs, prevent a false alarm, and be insensitive to the existence of bad data. Another problem is that the methods are sensitive to the existence of bad data.

### B. State Estimation with SCADA and PMUs

PMU measurements are a linear mapping of voltages while SCADA measurements are quadratic forms of voltages. An ideal state estimator should be capable of processing both measurements (not ignore higher order terms because the measurements may be taken at multiple states) and be efficient in computation for real-time estimation when an event is detected.

It is possible to apply the traditional methods when data from both SCADA and PMUs are available. Many research activities have been ongoing to integrate the data [8-14]. The most intuitive approach is to accept the voltages monitored by PMUs in the state estimation problem. This state estimation capability yields an efficient computation as the dimension of a data matrix shrinks proportional to the number of PMUs. When the voltages measured using PMUs are significantly different than those from the old state when both measurements were taken, all the errors are assigned to the remaining voltages in the variable space, which leads to an imprecise voltage estimate. An advanced method updates voltages when a new set of PMU data is available. The process is as follows: 1) estimate the voltages when data from SCADA and PMUs are available, and 2) linearly update the voltages when new sets of PMU data arrive between SCADA measurements. While this method yields an improved result over the intuitive approach does and yields a fast computation, it is difficult to update the voltages of the buses not monitored by PMUs. Therefore, this algorithm also suffers from the significant errors in the voltages at the buses. A shortcoming of both approaches is not combining the data; rather, they choose the results from the PMU measurements (intuitive method) or from the result of the traditional state estimation (linear update method). A hybrid method [15] is proposed to accommodate data from SCADA and PMUs. However, it was assumed that all the injections and voltage magnitudes of the buses that are not monitored by PMUs are available. In many situations where an event occurs, the assumptions do not hold. If the old values estimated are from the previous state estimation, the hybrid method results in a large error at the estimated voltage. In updating their sensitivity matrix, the estimate of the voltage magnitudes enters into the denominator, which makes the error propagation significantly to the solution. An ideal method should consider the original nonlinear power balance equation so the imprecise dispatch does not increase the error at iteration.

## IV. THEORIES

### A. Nonlinear Dynamic Tools – The Lyapunov Dimension and Exponent

Voltages are a good measure to represent the state of the system, and they depend on the generation dispatch and load profiles over the network. Therefore, the voltages indicate when to check if there is any change in the injections that can be affected by an event. Once the dispatch is determined, the voltages move toward an equilibrium. However, in real power system operation, loads and renewable energy resources are consistently changing, and therefore the voltages of a real system move toward any equilibrium. This reaction is aperiodic behavior, and the change is sensitive to the loading conditions and the generation. Therefore, it is impossible to perform a long-term prediction. Instead, they show aperiodic long-term behavior in a deterministic way that exhibits sensitive dependence on the loading conditions, termed chaos. A characteristic of chaos is self-similarity (i.e., fractal) [16].

The dimension of a fractal is the minimum number of independent variables to describe the fractal. Two different fractals may have the same dimension value, but multiple fractals with different values of dimension are guaranteed to be different (i.e., the value of dimension is a signature of a change in the state of the system of interest). This property fulfills a requirement of a monitoring system. In practical applications where the geometric object is reconstructed from a finite sample of data points with errors, correlation dimension is most widely used when it can be computed from a set of points on a phase space [17-19]. The measured quantity from PMUs is in time series, not on a phase space. Using the delay method, it is possible to reconstruct PMU data on a phase space. Once the fractal (voltages $v$) is reconstructed on a phase space $x$, the correlation dimension $d$ [16] is:

$$C(\varepsilon) = \frac{2}{N(N-1)} \sum_{i=1}^{N} \sum_{j=1+n_{\min}}^{N} H(\varepsilon - \|x_i - x_j\|) \propto \varepsilon^d \quad (1)$$

When a fractal undergoes a change, the Lyapunov exponent [20] is a quantity to measure the time response. The exponent shows how far two initially adjacent trajectories stay near. When a current state is analyzed, an estimate (initially close to the current state) to near future within the Lyapunov exponent is valid, i.e., $\|\delta(t)\| \propto \|\delta_0\| e^{\lambda t}$ [20]. The exponent can be positive, zero, or negative. Zero or negative exponents indicate that any data after a change occurs are all relevant in estimating the future state. For a positive exponent, two initially neighboring trajectories diverge as shown in Fig. 3.

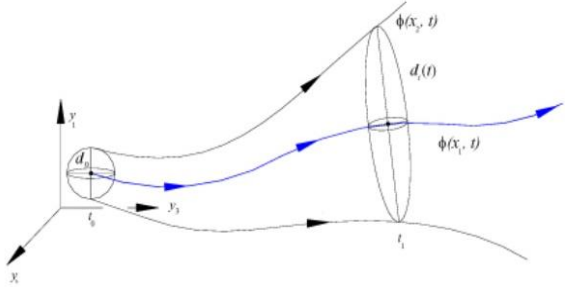

Figure 3. Schematic diagram of the Lyapunov exponent $\lambda$ that initially close states within $\delta_0$ stay close $\delta_0(t)$.

This nonlinear dynamics theory is applied to the electricity price time-series [21]. The time-series indicates how the state of electricity markets changes over time. For a controlled environment where multiple events are simulated, the prices from simulation are obtained. There are two events where a subset of the market participants changes their strategies permanently. Therefore, immediately after the events, the Lyapunov dimensions should undergo changes. Once the change in the dimension is detected, the Lyapunov exponents are evaluated (see Fig. 4). Between the first and second events, several "temporary changes" in their strategies are simulated. As a result, the event does not build up a permanent change, which should yield no change in dimension. Without detecting the change in dimensions, the tool does not check the exponent. However, for testing the possibility of a false alarm, the exponents are evaluated after the temporary changes and it was expected that there would be no changes in dimension and negative exponent. The results are illustrated in Fig. 4, which show no false alarm and, therefore, no capability to alarm.

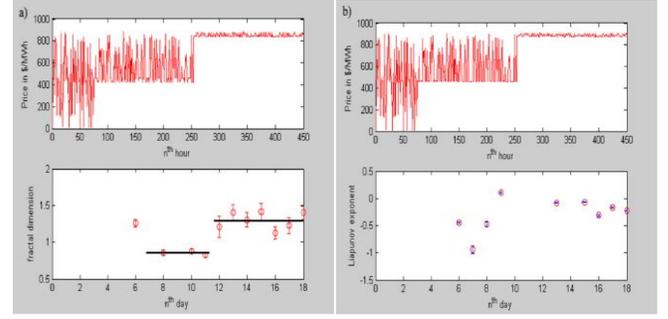

Figure 4. Daily check for a) dimension and b) the Lyapunov exponent calculated with the historical nodal price data obtained from the simulations [21].

*B. Real-time Monitoring Tool*

Situational awareness is achieved through the Lyapunov dimension and exponent, as listed under the situational awareness procedure. Over time, SCADA measurements are taken and the weight factor $W$ associated with the SCADA measurements decrease regardless of the situational awareness.

*Situational Awareness*

Set $\tau_{init} = 0$ ($\tau_{init}$ is the initial measurement time for investigation)
(1) Compute the Lyapunov dimension at time $t$ with time series data in [$\tau_{init}$, $t$]
(2) If the Lyapunov dimension is changed significantly, an event is alarmed and update $\tau_{init} = t$
Otherwise, go to Step (1)
(3) Compute the Lyapunov exponent with time series data in [$\tau_{init}$, $t$]
(4) If the Lyapunov exponent is negative or zero, reduce the weight factor of old SACDA measurements and perform a run state estimator with SCADA and PMU measurements
(5) Otherwise, update SCADA measurements. When no new SCADA measurements are available, reduce the weight factor of old SCADA data significantly. Then run a state estimator with the weighted old SCADA and new PMU measurements

The $j^{th}$ diagonal elements in $W$ reflect the error in the $j^{th}$ measurments $\delta$. The weight factor for SCADA measurements exponentially decreases as $t$ increases, i.e., $\|\delta(t)\| \propto \|\delta_0\| e^{\lambda t}$ [20] where $\|\delta_0\|$ is estimated by the state estimation with both PMU and SCADA measurements; $\lambda$ is the reciprocal Lyapunov exponent; and t is the time gap in the measurements of the most recent SCADA and the real-time PMU.

When a new set of SCADA measurements are available, old SCADA measurements are discarded for state estimation. The weight factor is updated as time goes from the measurement 1) minor decrease for Period 3, 2) moderate decrease for Period 2, or 3) significant decrease for Period 1 where Period 1, 2, and 3 are illustrated in Fig. 5.

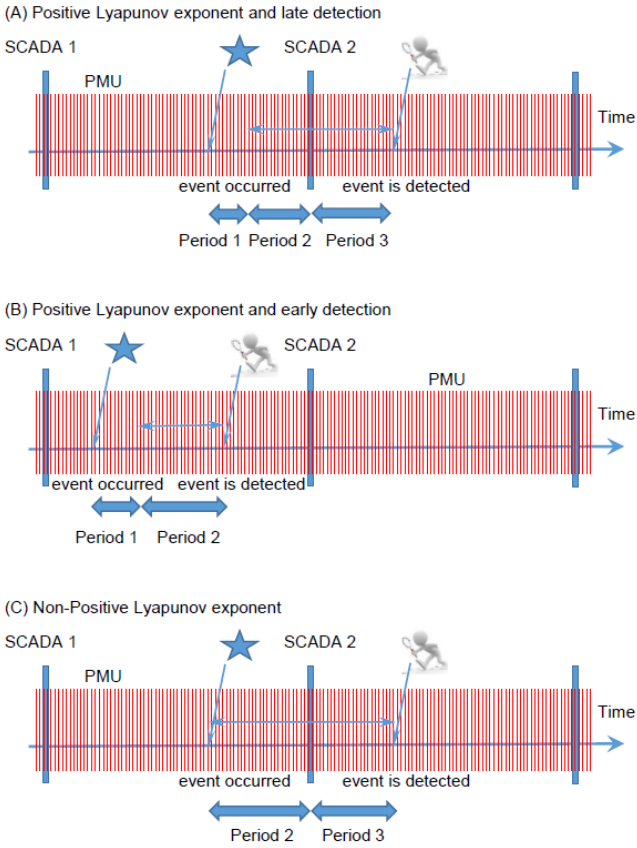

Figure 5. Situational awareness using PMU data while both data from SCADA and PMUs be used for state estimation. The double arrow indicates the fraction of data that are relevant to estimate the future state for the case: (A) Positive the Lyapunov exponent with delayed event detection, (B) Positive the Lyapunov exponent with early event detection, and (C) Non-positive the Lyapunov exponent.

### C. State Estimation using Kronecker Product

Power-balance equations are traditionally constructed in the polar coordinate system as voltages are phasor. Recently, this trend has been revisited and many researchers are formulating equations in the Cartesian coordinate system. When they are constructed in the Cartesian coordinate, they become quadratic equations at the $j^{th}$ bus from an $N$-bus system:

$$v^T B_j v = p_j \text{ and } v^T B_{j+N} v = p_{j+N} \quad (2)$$

where $v$ and $p$ are the voltage vector and a measurement that corresponds a quadratic equation in $v$, respectively. Quadratic function of voltages is a smoothly varying function, and Lipschitz continuous. The third or higher order terms in Taylor's expansion of the equations are zeros, which is a benefit when using the Cartesian coordinate system over the polar coordinate system.

In addition to the power balance equation, the voltage magnitude at the $j^{th}$ bus is also in the quadratic form similar to (2). Traditional power flow algorithms require two pieces of information at each bus so that four variables can be evaluated with two power balance equations. The required information depends on the type of buses. However, with the grid integration of new energy resources, it would be rather difficult to clearly define the type of buses. In general, $2N$ pieces of information are necessary to solve power-flow equations because there are $4N$ variables and $2N$ equations. It is not necessary to define 2 variables at each node. Data from SCADA measurements are all in the quadratic form as shown in (1). All the quadratic nodal equations are integrated into a single equation by using the Kronecker product:

$$v^T B_j v = p_j \rightarrow \begin{bmatrix} vec(B_1)^T \\ \vdots \\ vec(B_{2N})^T \end{bmatrix} v \otimes v = \begin{pmatrix} p_1 \\ \vdots \\ p_{2N} \end{pmatrix} \quad (3)$$

For the state estimation, the measurement vector and the data matrix have more than $2N$ rows and each measurement can have weight factor in (3), i.e., $\min_v \|W[\tilde{A}(v \otimes v) - \tilde{b}]\|_2$ [22]. Since the data from PMUs are linear in voltages, $l_j v = m_j$, they may not be expressed in terms of the quadratic expression in (3). However, they can be easily transformed into the quadratic equations, and the generic form of all equations for formulating a least square problem is:

$$\tilde{A} \begin{pmatrix} v \\ 1 \end{pmatrix} \otimes \begin{pmatrix} v \\ 1 \end{pmatrix} = \tilde{b} \quad (4)$$

where $\tilde{A} = \left[ vec\left(\frac{1}{2}\begin{bmatrix} 0 & l_1^T \\ l_1 & 0^{1\times 1} \end{bmatrix}\right) \cdots vec\left(\begin{bmatrix} B_{2N} & 0 \\ 0 & 0^{1\times 1} \end{bmatrix}\right) \right]^T$ and $\tilde{b} = (m_1 \cdots p_{2N})^T$.

An alternating least square (ALS) algorithm is employed to solve the problem. An ALS algorithm solves for the left $v$ first by setting the right $v$ at the value from the previous iteration, and then solves for the right $v$ by setting the left $v$ at the value from the left solution. The ALS algorithm suffers from 1) no convergence guarantee and 2) strong dependence of convergence on the initial guess to the solution. It is difficult to find a good initial point and the black triangle in Fig. 6 shows the performance of the ALS algorithm with a randomly chosen initial point. The solution the ALS finds is same when applying the Newton-Raphson method (local minimizer). Error minimization referred in Fig. 6 is $\left\| W\left[\tilde{A}\begin{pmatrix} v \\ 1 \end{pmatrix} \otimes \begin{pmatrix} v \\ 1 \end{pmatrix} - \tilde{b}\right] \right\|_2 / \|W\tilde{b}\|_2$.

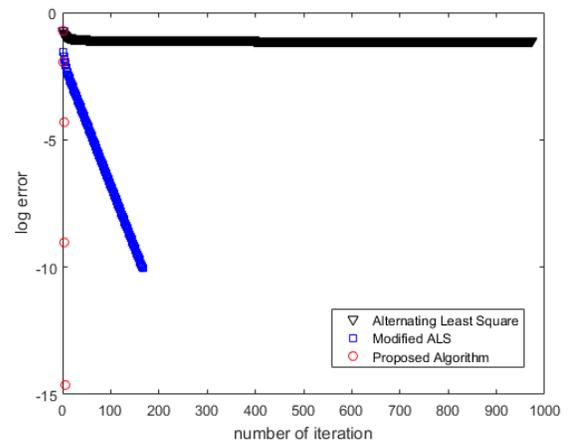

Figure 6. The comparison of the performance; ALS (triangle), Modified ALS (square), and Proposed algorithm (circle).

I modified the ALS algorithm by updating the average of the left and the right solutions. This simple step makes the computation highly efficient (see the blue square in Fig. 6 for the performance). However, it is not possible to prove the convergence of this modified ALS. For example, in the worst-case scenario, it may not find a solution. I propose a new algorithm to solve the least square problem

$$y^*_{k+1} = \arg\min_y \left( \|\varepsilon_k - A_k y\|_2 + \|W\tilde{A}\|_2 \|y\|_2^2 \right) \quad (5)$$

where $\varepsilon_k$ is the error in the previous iteration, i.e., $\varepsilon_k = W\tilde{b} - W\tilde{A}\begin{pmatrix}v_k\\1\end{pmatrix} \otimes \begin{pmatrix}v_k\\1\end{pmatrix}$ and $A_k = 2W\tilde{A}\left[I_{2N+1} \otimes \begin{pmatrix}v_k\\1\end{pmatrix}\right]\begin{bmatrix}I_{2N}\\0\end{bmatrix}$.

Since the data matrix $A_k$ is a full-column matrix and a part of a Tikhonov regularization term (the second term), the problem is convex. This unconstrained optimization is equivalent to the following constrained one:

$$y^*_{k+1} = \arg\min_y \|\varepsilon_k - A_k y\|_2$$

$$s.t. \|y\|_2 \leq \sqrt{\sum_{j=1}^{2N}\left[\frac{(e_j^T \Sigma_k e_j)(e_j^T U_k^T \varepsilon_k)}{(e_j^T \Sigma_k e_j)^2 + \|W\tilde{A}\|_2}\right]^2} \quad (6)$$

where $A_k = U_k \Sigma_k V_k^T$, and $\varepsilon$ is in the cardinality of $m$ ($m > 2N$, the number of measurements). Algorithm I reduces error at every iteration:

$$\|\varepsilon_{k+1}\|_2 = \|\varepsilon_k - A_k y^*_{k+1} - W\tilde{A} y^*_{k+1} \otimes y^*_{k+1}\|_2$$

$$\leq \|\varepsilon_k - A_k y^*_{k+1}\|_2 + \|W\tilde{A}\|_2 \|y^*_{k+1}\|_2^2 \leq \|\varepsilon_k\|_2$$

Equality holds only when $y^*_{k+1}$ equals zero, i.e., when the algorithm converges. The convergence rate of this algorithm is proven quadratic, and β is found that

$$\|J(v_k)^{-1}\|_2 = \left\|\left(A_k^T A_k + \|W\tilde{A}\|_2 I\right)^{-1}\right\|_2$$
$$= \left\|\left(\Sigma_k^T \Sigma_k + \|W\tilde{A}\|_2 I\right)^{-1}\right\|_2 < \frac{1}{\|W\tilde{A}\|_2} = \beta \quad (7)$$

The inequality holds because $A_k$ is a full-column-rank matrix. An analytic solution exists for the optimization in (4) [5]:

$$y^*_{k+1} = \sum_{j=1}^{2N}\left[\frac{(e_j^T \Sigma_k e_j)(e_j^T U_k^T \varepsilon_k)}{(e_j^T \Sigma_k e_j)^2 + \|W\tilde{A}\|_2}\right](V_k e_j) \quad (8)$$

(8) indicates that, even though $\Sigma_k$ includes zero (or near zero) singular values in its diagonal which mean $A_k$ is ill-conditioned, $y^*_{k+1}$ is still well-defined because strictly positive element ($\|W\tilde{A}\|_2$) makes the denominator strictly positive. Therefore, different from Newton-Raphson process, the ill-conditioned data matrix does not result in numerical instability. This update process is not based upon linear approximation, rather its update is from a convex optimization with a Tikhonov regularization term. Therefore, a curvilinear update is performed at iteration. Algorithm I, new tensor-based state estimation method proposed here, yields a quadratic convergence. This method employs the analytic solution of least square minimization over a sphere once a singular value decomposition (SVD) is performed of a data matrix at iteration [4]. This proposed algorithm does not involve any approximation and the error reduces at every iteration, regardless of the choice of an initial guess.

### *Algorithm I*

Set $k = 0$ ($k$ is the iteration index) and $\varepsilon_{th}$ (tolerance in error)

(1) If $\left\|W\tilde{b} - W\tilde{A}\begin{pmatrix}v_k\\1\end{pmatrix} \otimes \begin{pmatrix}v_k\\1\end{pmatrix}\right\|_2 \leq \varepsilon_{th}$, terminate the process

(2) Otherwise solve (6) for $y^*_{k+1}$, i.e., $\left(A_k^T A_k + \|W\tilde{A}\|_2 I\right) y^*_{k+1} = A_k^T \varepsilon_k$

(3) Update $v_{k+1} = v_k + y^*_{k+1}$; set $k$ with $k+1$; and go to (1)

As shown in Fig. 6, it is clear that the Kronecker least square approach outperforms ALS methods in terms of finding a proper solution, keeping track of the change occurring on the system, and computational efficiency both in the quality of the solution and the less number of iteration to get to the solution. Therefore, the proposed Kronecker least square approach is most suitable for precise state estimation. It is also worthwhile to mention that the traditional Newton-Raphson method does not converge since the higher order terms are too large to ignore, which results in divergence.

### D. Proof of Convergence of the Proposed Algorithm

The local convergence of the proposed algorithm is sketched below. Multiple initial points were tested to see if they result in a different solution, and they return all the same numerical solution. It was concluded that Algorithm I seems to show a global convergence numerically.

### *Convergence theorem*

**Assumptions**: There exists $v_\infty \in R^n$; an open convex set $D$; and positive scalar $r > 0$ such that $\mathcal{M}(v_\infty, r) \subset D$ where $\mathcal{M}(v_\infty, r) = \{v \in R^n : \|v - v_\infty\| < r\}$

**Observations**:
1) Given F: $R^n \to R^n$, $F(v_k) = -A_k^T \varepsilon_k$ is continuously differentiable in an open convex set $D \subset R^n$; and
2) Given J: $R^n \to R^n$, $J(v_k) = A_k^T A_k + \|W\tilde{A}\|_2 I$ is Lipschitz continuous in $\mathcal{M}(v_\infty, r)$; $\|J(v_k)^{-1}\|_2 \leq \beta$; $v_0 \in R^n$; and
3) Algorithm I at each iteration solves $J(v_k) y_{k+1} = -F(v_k)$ and update $v_{k+1} = v_k + y^*_{k+1}$.

**Results**: There exists $\Delta > 0$ such that for all $v_0 \in \mathcal{M}(v_\infty, \Delta)$ the sequence $v_1, v_2, \ldots$ generated by $v_{k+1} = v_k - J(v_k)^{-1} F(v_k)$ is well defined regardless the condition number of $A_k$, converges to $v_\infty$, and obeys $\|v_{k+1} - v_\infty\| \leq M \|v_k - v_\infty\|^2$ where $M$ is a finite number. Detailed proof is found in Ref. [23].

### E. Initial Point Dependence and Extension to General Polynomial Functions

It is recognized that the convergence of Newton-Raphson method can strongly depend upon the choice of an initial point.

Suppose we aim to find roots of $f(x) = x^3 - 5x$, and the initial guess is 1. Newton-Raphson process will oscillate between +1 and -1 with no progress, i.e., fails to find a solution. The proposed approach can be extended to incorporate any polynomial function. Here is an extension setup:

$$x^3 - 5x = 0 \Leftrightarrow x^2 - y = 0, \ xy - 5x = 0$$
$$\rightarrow z^T A_1 z = 0, \ z^T A_2 z = 0, \ z^T A_3 z = 1$$

where $A_1 = \begin{bmatrix} 1 & 0 & 0 \\ 0 & 0 & -\frac{1}{2} \\ 0 & -\frac{1}{2} & 0 \end{bmatrix}, A_2 = \begin{bmatrix} 0 & \frac{1}{2} & -\frac{5}{2} \\ \frac{1}{2} & 0 & 0 \\ -\frac{5}{2} & 0 & 0 \end{bmatrix},$

$A_3 = \begin{bmatrix} 0 & 0 & 0 \\ 0 & 0 & 0 \\ 0 & 0 & 1 \end{bmatrix}, \text{ and } z = \begin{pmatrix} x \\ y \\ 1 \end{pmatrix}$ (9)

With a starting point of $x_0 = 1$, the performances of Newton-Raphson method and of the proposed method are illustrated in Fig. 7, and it clearly indicate that the proposed method converges when the conventional Newton-Raphson process fails to do so.

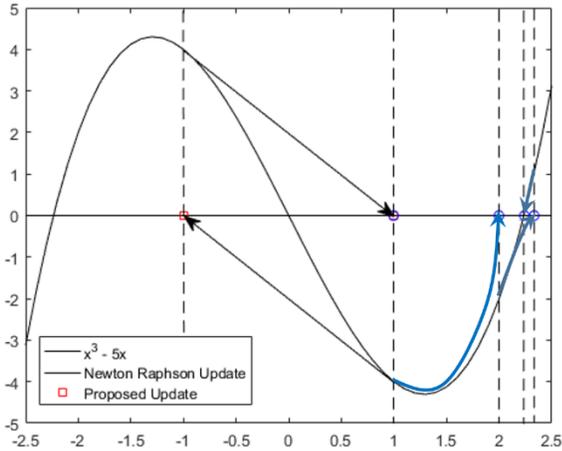

Figure 7. Comparison in the convergences of Newton-Raphson and proposed methods.

The convergence is achieved when the initial guess of $y_0$ is greater than or equal to 3.5. When a smaller value was chosen, the proposed algorithm does converge even though it finds 0 instead of $\sqrt{5}$. Due to the curvilinear update, the number of iteration is lower than that of the linear update such as Newton-Raphson method.

## V. SIMULATION RESULTS

### A. Situational Awareness

A set of continuous power flow simulation is performed on the IEEE 30 bus case; the voltages at Bus 12 are monitored. Since the voltage magnitudes are close to 1 per unit and the voltage angles are close to zero, the real part of the voltages is near unity. Therefore, the imaginary part of the voltage is monitored. The load increases slowly and has a periodic cycle. Between the 2nd and the 14th seconds, a wind turbine at Bus 4 injects more power than expected.

As shown at the top plot in Fig. 8, significant changes in the dimension are detected after the changes at both the 2nd and the 14th seconds, which indicates the changes in the system – situational awareness. The Lyapunov exponent at the bottom plot of Fig. 8 shows how recent data are relevant to estimate the current states. Except two points immediately after each change, all the exponents are non-positive, i.e., all the previous data are relevant. At the two points after the changes, the exponents are positive, i.e., only $1/\tau_L$ recent data are relevant. $\tau_L$ are 0.17 and 0.08, which means data measured 5.88 and 12.5 seconds before the PMU measurements are relevant to estimate the state.

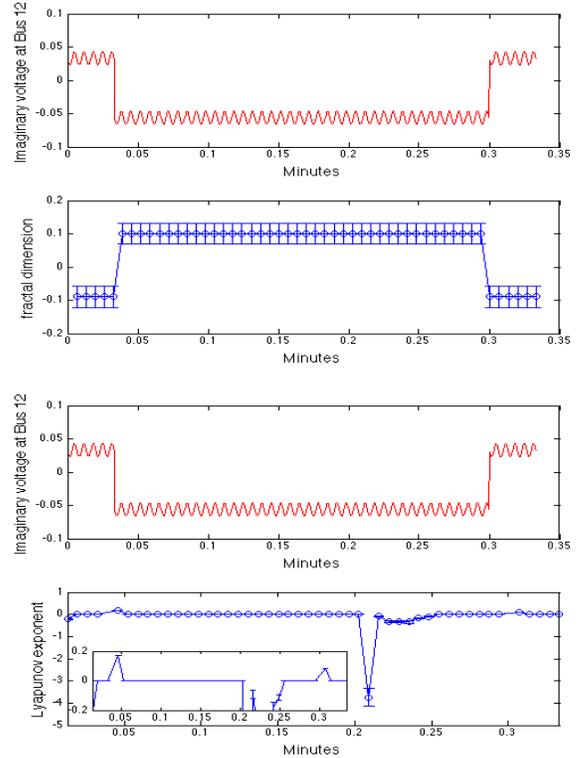

Figure 8. Lyapunov dimension (top) and exponent (bottom) from the imaginary voltage at Bus 12.

Dimension and Lyapunov exponent provide information on 1) the change in the state of the system, and 2) the fraction of data is relevant to estimate the state. From the results, the nonlinear dynamic tool is useful for situational awareness in that it properly detects the change in the state and does not show a false alarm.

To compare the performances of the proposed algorithm to those of other methods, a set of simulations is performed, wind output changes rapidly, and PMUs track the voltages at 30 Hz while SCADA takes measurements every 2 seconds. The linear method finds practically no changes in the voltages, but the hybrid and the proposed methods detect the changes as shown in Fig. 9. PMU measurement in the x-axis in Fig. 8 is equivalent to time $t$ from the last SCADA measurement. The change of voltage refer $\|v_{est} - v_0\|_2 / \|v_{ref} - v_0\|_2$ where $v_{est}$, $v_0$, and $v_{ref}$ refer estimated voltage, voltage at $t = 0$, and reference voltages that real parts are all unities and imaginary parts are all zeros, i.e., $v_{ref} = [1_N; 0_N]$, respectively. The weight factor $W$ is estimated based upon the Lyapunov exponent using $\|\delta(t)\| \propto \|\delta_0\| e^{\lambda t}$.

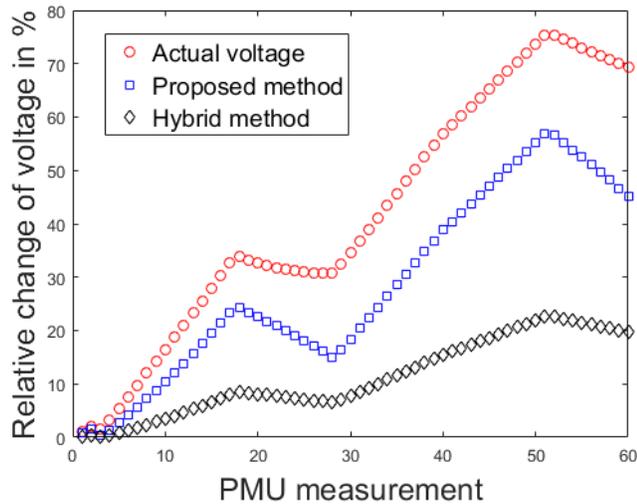

Figure 9. Comparison in the performances of hybrid and proposed methods.

## VI. CONCLUSIONS

PMUs are integrated to enhance the situational awareness of large scale power networks. Due to the high integration costs of PMUs, it is not practical to equip PMUs at all the nodes in the network. Data from SCADA would be ideal to complement the geographically limited coverage of PMUs. Current situational awareness technology does not incorporate date from PMU and SCADA. A new state estimation method along with nonlinear monitoring capability is proposed to simultaneously integrate PMU and SCADA measurement. A weighted least square approach based on a tensor computation, Kroncker product, is tested, and shows an excellent performance in terms of the quality of the solution and the computational efficiency. The proposed method can be extended to any equations with polynomial functions.

## VII. ACKNOWLEDGEMENTS

We thank Professor Charles Van Loan at Cornell University for the suggestion in the tensor formulation for the state estimation; for the introduction of the alternating least square approach and of the preconditioning; and for the insights and the discussion in numerical computation.